%% file: main.tex
\documentclass[10pt]{IEEEtran}

\input{./include/header}

\input{./include/macros}
\loadglsentries{./include/acronyms}
\glsenableentrycount

\begin{document}
 
\title{Spectral CT Two-step and One-step Material Decomposition using Diffusion Posterior Sampling}

\author{\IEEEauthorblockN{Corentin Vazia\IEEEauthorrefmark{1}, Alexandre Bousse\IEEEauthorrefmark{2}, Jacques Froment\IEEEauthorrefmark{1}, Béatrice Vedel\IEEEauthorrefmark{1}, Franck Vermet\IEEEauthorrefmark{3}, Zhihan Wang\IEEEauthorrefmark{2}\IEEEauthorrefmark{3}, Thore Dassow\IEEEauthorrefmark{2}\IEEEauthorrefmark{5}, Jean-Pierre Tasu\IEEEauthorrefmark{4} and Dimitris Visvikis\IEEEauthorrefmark{2}}

\IEEEauthorblockA{\IEEEauthorrefmark{1}Univ Bretagne Sud, CNRS 6205, LMBA, F-56000 Vannes, France.}

\IEEEauthorblockA{\IEEEauthorrefmark{2}LaTIM  U1101, Université de Bretagne Occidentale, Brest, France.}

\IEEEauthorblockA{\IEEEauthorrefmark{3}Univ Brest, CNRS, UMR 6205, Laboratoire de Mathématiques de Bretagne Atlantique, France.}

\IEEEauthorblockA{\IEEEauthorrefmark{4}Departement of Radiology, University Hospital Poitiers, France.}

\IEEEauthorblockA{\IEEEauthorrefmark{5}Siemens Healthcare SAS, Courbevoie, France.}
}

\maketitle

\input{./content/abstract}
\glsresetall
\input{./content/intro}
\input{./content/method}

\input{./content/results}

\input{./content/conclusion}

\section{Acknowledgments}
\label{sec:acknowledgments}
This work was conducted within the France 2030 framework programmes, Centre Henri Lebesgue ANR-11-LABX-0020-01 and French National Research Agency (ANR) under grant No ANR-20-CE45-0020, France Life Imaging under grant No ANR-11-INBS-0006, and with the support of \textit{Région Bretagne}. 

\bibliographystyle{IEEEtran}
\bibliography{IEEEabrv,biblio}

\end{document}

%% file: include/header.tex
\usepackage{graphicx}

\usepackage{bm}

\usepackage[acronym]{glossaries}

\usepackage{epstopdf}
\usepackage{cite}

\usepackage{tikz}
\usepackage{tikzscale}
\usetikzlibrary{arrows,shapes,calc,decorations.pathreplacing,intersections,quotes,angles,positioning,fit,petri,through,shadows,plotmarks,spy}
\usepackage{pgfplots}
\usepackage{import}
\usetikzlibrary{positioning}
\usetikzlibrary{3d}

\usepackage{amssymb}
\usepackage{microtype}

\usepackage{overpic}
\usepackage{breqn}

\usepackage{svg}
\usepackage{calc}

\usepackage{caption}
\usepackage{subcaption}
\usepackage{array}

%% file: include/macros.tex
\newlength{\tempdima}
\newcommand{\rowname}[1]
{\rotatebox{90}{\makebox[\tempdima][c]{\scriptsize#1}}}
\newcommand{\argmin}{\operatornamewithlimits{arg\,min}}
\newcommand{\argmax}{\operatornamewithlimits{arg\,max}}

%% file: include/acronyms.tex
\newacronym{ct}{CT}{computed tomography}
\newacronym{pcct}{PCCT}{photon-counting CT}
\newacronym{sct}{sCT}{spectral CT}
\newacronym{lac}{LAC}{linear attenuation coefficient}
\newacronym{tv}{TV}{total variation}
\newacronym{tnv}{TNV}{total nuclear variation}
\newacronym{jtv}{JTV}{joint total variation}
\newacronym{sde}{SDE}{stochastic differential equation}
\newacronym{fbp}{FBP}{filtered back projection}
\newacronym{odps}{ODPS}{\textit{One-step Diffusion Posterior Sampling}}
\newacronym{tdps}{TDPS}{\textit{Two-step Diffusion Posterior Sampling}}
\newacronym{dtv}{DTV}{directional total variation}
\newacronym{psnr}{PSNR}{peak signal-to-noise ratio}
\newacronym{ssim}{SSIM}{structural similarity}
\newacronym{pnll}{PNLL}{penalized negative log-likelihood}
\newacronym{map}{MAP}{maximum \emph{a~posteriori}}
\newacronym{pdf}{PDF}{probability distribution function}
\newacronym{nn}{NN}{neural network}
\newacronym{cnn}{CNN}{convolutional neural network}
\newacronym{dps}{DPS}{diffusion posterior sampling}
\newacronym{dm}{DM}{diffusion model}
\newacronym{mbir}{MBIR}{model-based iterative reconstruction}
\newacronym{mse}{MSE}{mean squared error}
\newacronym{2d}{2-D}{two-dimensional}
\newacronym{3d}{3-D}{three-dimensional}
\newacronym{ddpm}{DDPM}{denoising diffusion probabilistic models}
\newacronym{nist}{NIST}{National Institute of Standards and Technology}
\newacronym{wls}{WLS}{weighted least-squares}
\newacronym{pwls}{PWLS}{penalized weighted least-squares}

%% file: content/abstract.tex
\begin{abstract}
	This paper proposes a novel approach to spectral \gls{ct} material decomposition that uses the recent advances in generative \glspl{dm} for inverse problems.  Spectral \gls{ct} and more particularly \gls{pcct} can perform transmission measurements at different energy levels which can be used for material decomposition.
	It is an ill-posed inverse problem and therefore requires regularization. \Glspl{dm} are a class of generative model that can be used to solve inverse problems via \gls{dps}.

	In this paper we adapt \gls{dps} for material decomposition in a \gls{pcct} setting. We propose two approaches, namely \gls{tdps} and \gls{odps}. Early results from an experiment with simulated low-dose \gls{pcct} suggest that \glspl{dps} have the potential to outperform state-of-the-art \gls{mbir}. Moreover, our results indicate that \gls{tdps} produces material images with better \gls{psnr} than images produced with \gls{odps} with similar \gls{ssim}.

\end{abstract}

%% file: content/intro.tex
\section{Introduction}

Spectral \gls{ct} and the energy-dependent attenuation allow to reconstruct images of the materials  present within the scanned object or patient \cite{Alvarez_1976}. It is an ill-posed inverse problem and requires regularization, or prior. Conventional \gls{mbir} techniques typically include two types of approaches, namely two-step and one-step techniques.

On one hand, two-step techniques aim at reconstructing high-quality multi-energy images which are then  used for material decomposition. The images can be reconstructed synergistically to leverage the information shared across channels, for example by enforcing structural similarities \cite{Rigie_2015}, low-rank  \cite{gao2011robust,he2023spectral} or similarities with a reference clean image \cite{Synergistic_DTV}. Alternatively, the regularizer can also be trained \cite{zhang2016tensor,perelli2022multi,Uconnect} (see \cite{Review_spectral_ct} for a review).

On the other hand, one-step techniques directly estimate the material images from the raw projection data. This is also achieved with the help of \gls{mbir} techniques to minimize the negative log-likelihood which preserves the Poisson statistics. Likewise, these methods are regularized, for example by promoting pixel-wise material separation  \cite{gondzio2022material}.

Generative models and particularly \glspl{dm} have shown promising results for sampling realistic images from a training dataset \cite{dhariwal2021diffusion}. More recently, they have been used to solve inverse problems by guiding the sampling scheme \cite{song2021scorebased} or via \gls{dps} \cite{chung2023diffusion}. 
 
In this work, we propose to solve the material decomposition inverse problem by \gls{dps}.

We propose two approaches, \gls{tdps}, based on our previous work \cite{vazia2024spectral}, and \gls{odps}, for the decomposition of bone and soft tissue materials.

Section~\ref{section:method} presents our paradigm, starting from the forward model and the two approaches (two-step and one-step) in Section~\ref{section:forward_models}, followed by the description of the two approaches we developed, namely \gls{odps} and \gls{tdps} in Section~\ref{section:DPS}. Section \ref{section:res} presents results of two-material decomposition in simulated \gls{pcct} with different X-ray photon flux and compares the two methods proposed to other decomposition techniques. Finally, we discuss and conclude our work in Section~\ref{section:conc}.

%% file: content/method.tex
\section{Method}\label{section:method}

\subsection{Spectral CT and Material Decomposition} \label{section:forward_models}

Building on advances on X-ray \gls{ct}, it is possible to leverage the energy dependence of the \gls{lac} in order to reconstruct multiple images of the same scanned object but at different level of energy $E$.

We (temporarily) denote by
\begin{equation}\label{eq:attn}
 \bm{X}(E) = [X_1(E), X_2(E), \dots, X_J(E)] \in \mathbb{R}^J	\nonumber
\end{equation}
the energy-dependent attenuation image (random vector) where $J$ is the number of pixels and $X_j(E)$  the \gls{lac}  at pixel $j$ and energy $E$.
 
For each pixel $j$, $X_j$ can be decomposed as a sum over the materials composing the scanned object or patient. Denoting by $Z_{j,n}$ the $n$-th material concentration at pixel location $j$, we have the following material composition

\begin{align}
	X_j(E) = {} & \sum_{n=1}^{N} f_n(E) Z_{j,n} \nonumber \\
	\triangleq {}&	 \mathcal{F}(\bm{Z}_j,E)  \label{eq:decomp}
\end{align}
where $f_n(E)$ is the known $n$-th material attenuation function multiplied by the density of the corresponding material, $\bm{Z}_j = \{Z_{j,n}\}_{n=1}^N \in \mathbb{R}^N$ and $N$ is the number of materials. The $\bm{Z}_j$'s are regrouped into a material image $\bm{Z} = \{Z_j\}_{j=1}^J \in \mathbb{R}^{J\cdot N}$, and $\mathcal{F}$ in \eqref{eq:decomp} can be generalized to the entire image as 
\begin{align}
	\mathcal{F}(\bm{Z},E) = {} & \bm{X}(E) \nonumber \\
	\triangleq  {} & \{ \mathcal{F}(\bm{Z}_j,E) \}_{j=1}^J \in \mathbb{R}^J  \, .  \label{eq:Fcont}
\end{align}

We now consider the standard setting of \gls{pcct}. The energy spectra of the X-ray beams is discretized into $K$ energy bins of the form $[E_k, E_{k+1}]$ and we denote by $Y_{i,k}$ the measurement for the energy bin $k$ and along the $i$-th ray, $i=1,\dots,I$.  The $Y_{i,k}$'s are random variables with conditional distribution
\begin{align}
	(Y_{i,k} \mid \bm{Z} = \bm{z} ) \sim {} &  	(Y_{i,k} \mid \bm{X}(\cdot) =   \mathcal{F}(\bm{z},\cdot) ) \nonumber\\
	\sim	{}  & \mathrm{Poisson} \left( \bar{Y}_{i,k} \left( \mathcal{F}(\bm{z},\cdot)   \right)	\right)  \nonumber
\end{align}
where the mean number of detection $\bar{Y}_{i,k}$ given some energy-dependent image $\bm{x}(E) = \mathcal{F}(\bm{z},E) $ is

\begin{equation}\label{eq:forward_cont}
	\bar{Y}_{i,k} \left( \bm{x}\right) \triangleq  \int h_{i,k} (E) \cdot \mathrm{e}^{-[\bm{\mathcal{A}}(\bm{x}(E))]_i} \, \mathrm{d}E \, ,
\end{equation}
$\bm{\mathcal{A}} \colon \mathbb{R}^J \to \mathbb{R}^{I}$ being the forward \gls{ct} operator and $h_{i,k}(E)$ being the photon flux for the energy bin $k$. The random variables $Y_{j,k}$ are conditionally independent  given $\bm{X}(\cdot)$ (and therefore given $\bm{Z}$). We regroup the measurement at bin $k$ into a random vector $\bm{Y}_k \in \mathbb{R}^I$ and the complete measurement into one random vector $\bm{Y} = \{\bm{Y}_k\}_{k=1}^K \in\mathbb{R}^{I\cdot K}$. 

In this work we consider the standard simplified model where the energy-dependent attenuation $\bm{X}(E)$ is ``energy-discretized'' into $K$ images $\bm{X}_k = [X_{1,k},\dots,X_{J,k}]\in \mathbb{R}^{J}$, one for each energy bin $k$, such that $\bm{X}_k$ corresponds to an average attenuation image for energy bin $k$, and we redefine $\bm{X} \triangleq \{\bm{X}_k\}_{k=1}^K \in \mathbb{R}^{J\cdot K}$ as the vector-valued image regrouping the $K$ energy bins. Moreover, we use an energy-discretized version of \eqref{eq:decomp}
\begin{align}\label{eq:F_discr}
	X_{j,k} = {} &  \sum_{n=1}^{N} f_{n,k} Z_{n,j} \nonumber \\
	\triangleq {} & \mathcal{F}_k(\bm{Z}_j)    \nonumber
\end{align}    
and we redefine the operator $\mathcal{F}$, initially defined in \eqref{eq:Fcont}, as the generalized material decomposition operator over the entire image for all energy bins:
\begin{equation}
	\mathcal{F}(\bm{Z}) \triangleq \{\mathcal{F}_k (\bm{Z})\}_{j,k=1}^{J,K} \in \mathbb{R}^{J\cdot K}   \, .  \nonumber 
\end{equation}
Finally, we have the following simplified forward model:
\begin{align} 
	(Y_{i,k} \mid \bm{Z} = \bm{z} ) \sim {} &  	(Y_{i,k} \mid \bm{X}_k = \mathcal{F}_k(\bm{z}) ) \nonumber \\
	\sim	{}  & \mathrm{Poisson} \left( \bar{Y}_{i,k} \left( \mathcal{F}_k(\bm{z}) \right)   \right)	\label{eq:poisson_disc}
\end{align}
where for some attenuation image $\bm{x}_k$ at bin $k$ the expected number of counts $\bar{Y}_{i,k}$ is given by a simplified version of \eqref{eq:forward_cont}: 
\begin{equation}\label{eq:forward_disc}
	\bar{Y}_{i,k} \left( \bm{x}_k\right) \triangleq  \bar{h}_{i,k}\cdot \mathrm{e}^{{-[\bm{\mathcal{A}}(\bm{x}_k)]_i}}
\end{equation}
with $\bar{h}_{i,k} = \int h_{i,k}(E)\mathrm{d}E$.

Thus, given an energy-binned measurement $\bm{Y} = \bm{y} \in \mathbb{R}^{I\cdot K}$,  \gls{map} spectral \gls{ct} material decomposition can be achieved in two ways: (i) the two-step approach, i.e., 
\begin{equation}\label{eq:2step}
	\hat{\bm{x}} \in \argmax_{\bm{x} \in \mathbb{R}^{I\cdot K}} \, p_{\bm{Y}|\bm{X} = \bm{x}}(\bm{y})\cdot  p_{\bm{X}}(\bm{x}) \quad \text{then solving} \quad \mathcal{F}(\bm{z}) = \hat{\bm{x}} 
\end{equation} 
and (ii) the one-step approach, i.e.,
\begin{equation}\label{eq:1step}  
	\hat{\bm{z}} \in \argmax_{\bm{z}\in \mathbb{R}^{J\cdot N} } \, p_{\bm{Y}|\bm{Z} = \bm{z}}( \bm{y}  ) \cdot  p_{\bm{Z}}(\bm{z}) 
\end{equation}
where $p_{\bm{Y}|\bm{X} = \bm{x}}$ and $p_{\bm{Y}|\bm{Z} = \bm{z}}$ are given by \eqref{eq:poisson_disc} 
and $p_{\bm{X}}$ and $p_{\bm{Z}}$ are respectively the prior \glspl{pdf} of $\bm{X}$ and $\bm{Z}$. 

Solving \eqref{eq:2step} and \eqref{eq:1step} is usually achieved with the help of \gls{mbir} techniques. In the case of the two-step decomposition \eqref{eq:2step}, the pseudo-inverse of $\mathcal{F}$, denoted $\mathcal{F}^\dag$, can be used to obtain $\hat{\bm{z}} = \mathcal{F}^\dag \left(\hat{\bm{x}}\right)$. The  log-priors $\log p_{\bm{X}}$ and $\log p_{\bm{Z}}$ are in general unknowns and need to be replaced by handcrafted regularizers. Examples of such regularizers for $\bm{x}\in\mathbb{R}^{J\cdot K}$ in the two-step approach \eqref{eq:2step}  include \gls{tv} or \gls{tnv} \cite{Rigie_2015} to promote structural similarities across channels or with a reference image \cite{Synergistic_DTV}, as well as low-rank regularizers \cite{gao2011robust,he2023spectral}. The regularizers  can also be trained, for example  with tensor dictionary learning \cite{zhang2016tensor}, convolutional dictionary learning \cite{perelli2022multi}  or U-Nets \cite{Uconnect}. Regularizers for material images $\bm{z} \in \mathbb{R}^{J\cdot N}$ can for instance promote neighboring pixels to have similar values while preserving edges \cite{Weidinger2016} or promote pixel-wise material separation \cite{gondzio2022material}.

\subsection{Diffusion Models} \label{section:DPS}

\Glspl{dm} \cite{ho2020denoising, song2020generative} are a new state-of-the-art \gls{cnn}-based generative models. In previous work \cite{vazia2024spectral} we proposed a \gls{dps} framework to sample the multi-energy image $\bm{X}\in \mathbb{R}^{J\cdot K}$. This method enables sampling images according to the joint \gls{pdf} of all channels simultaneously, leveraging inter-channel information. Compared to individually sampling each $\bm{X}_k$, this approach enhances image quality, thus potentially improving the quality of material images obtained within a two-step framework, namely \gls{tdps}. Similarly, \gls{dps} can be used in a one-step framework, namely \gls{odps}, to directly sample the material image $\bm{Z}\in \mathbb{R}^{J\cdot N}$ without reconstructing $\bm{X}$.  

\subsubsection{Image Generation}

We denote by $\bm{W} \in \{\bm{X},\bm{Z}\}$ the random vector which can be either $\bm{X}$ or $\bm{Z}$ depending on which problem we wish to solve (\gls{tdps} \eqref{eq:2step} or \gls{odps} \eqref{eq:1step}). In the following paragraph we briefly describe \glspl{dm} to sample $\bm{W}=\bm{w}$ from $p_{\bm{W}}$ and then how to leverage such models to sample from $p_{\bm{W} | \bm{Y}=\bm{y}}$.

Generative models are used to generate new samples from $p_{\bm{W}}$ trained from limited training dataset with empirical \gls{pdf} $p_{\mathrm{data}}$ that approximates $p_{\bm{W}}$. \Glspl{dm} have been recently introduced in image processing and have shown promising performances \cite{dhariwal2021diffusion}. 
Song et al. \cite{song2021scorebased} showed that \glspl{dm} can be viewed as a \gls{sde} framework. The general idea consists in using a diffusion \gls{sde} that pushes the initial distribution $p_0=p_{\bm{W}}$ into a white noise. The ``variance preserving''  forward \gls{sde} is (in the ideal case $\bm{W}_0 \sim p_{\bm{W}}$) \cite{ho2020denoising, song2021scorebased}
\begin{equation} \label{forward_sde}
	\mathrm{d}\bm{W}_t =-\frac{1}{2} \beta(t) \bm{W}_t\,\mathrm{d}t + \sqrt{\beta(t)}\mathrm{d}\bm{B}_t  \textrm{~for~} t \in [0,T] \nonumber
\end{equation}
where $\bm{B}_t$ is a standard Wiener process. The function $\beta \colon \mathbb{R} \to \mathbb{R}$ is chosen such that $\bm{W}_T$ approximately follows a standard normal distribution. We assume that for each $t$ in $[0,T]$,   $\bm{W}_t$ follows the \gls{pdf} $p_t$. According to Anderson \cite{anderson1982}, the corresponding reverse time \gls{sde} is
\begin{dmath} \label{reverse_sde}
	\mathrm{d}\bm{W}_t = \left[ -\frac{1}{2} \beta(t) \bm{W}_t - \frac{1}{2}\beta(t) \nabla (\log p_t)(\bm{W}_t) \right]\mathrm{d}t + \sqrt{\beta(t)}\mathrm{d}\bm{B}_t \, .
\end{dmath}
The term $\nabla (\log p_t)(\cdot)$ is called the score function. It is intractable and therefore we use a deep \gls{nn} $\bm{s}_{\bm{\theta}}(\bm{w}, t)$ parameterized by $\bm{\theta}$ in order to approximate it. Training $\bm{s}_{\bm{\theta}}$ with a \gls{mse} loss could be achieved as
\begin{equation}\label{eq:score_matching1}
	\hat{\bm{\theta}} \in \argmin_{\bm{\theta}}  \: \mathbb{E}_{t, \bm{W}_t} \left[ \lVert \bm{s}_{\bm{\theta}}(\bm{W}_t, t) - \nabla (\log p_t)(\bm{W}_t) \rVert_2^2 \right] \, \nonumber
\end{equation}	
where the expectation is taken with $t\sim \mathcal{U}\{0,T\}$ and $\bm{W}_t \sim p_{t}$. However, since $p_t$ is unknown, we use the following surrogate optimization problem  which leads to the same minimizer    \cite{Score2005}:
\begin{equation}\label{eq:score_matching2}
	    \hat{\bm{\theta}} \in \argmin_{\bm{\theta}} \: \mathbb{E}_{t, \bm{W}_0, \bm{W}_t | \bm{W}_0} \left[ \lVert \bm{s}_{\bm{\theta}}(\bm{W}_t, t) - \nabla (\log p_{t|0})(\bm{W}_t)  \rVert_2^2 \right]  \nonumber
\end{equation}	
where $\bm{W}_t | \bm{W}_0\sim p_{t|0}$ and  $p_{\mathrm{data}}$ is used instead of $p_0$ to compute the expectation. See \cite{song2021scorebased} (appendix C) for more details on the computation of $p_{t|0}$. In this work we use the denoising diffusion probabilistic model implementation \cite{ho2020denoising} where the \gls{nn} predicts noise added during the diffusion instead of the score. This is based on Tweedie's formula, which enables us to establish a connection between both aspects. Starting from white noise and following \eqref{reverse_sde} from $t=T$ to $t=0$ with the score replaced by $\bm{s}_{\hat{\bm{\theta}}}$, we obtain a realization of $\bm{W} \sim p_{\bm{W}} = p_0$.

\subsubsection{Solving Inverse Problems}\label{sec:inv_prob}

It is possible to leverage the generative capability of a \gls{dm} to regularize an inverse problem, see for instance \cite{song2022solving, chung2022improving}. The idea is to condition the reverse \gls{sde} \eqref{reverse_sde} on the measurements $\bm{Y}=\bm{y}$. This leads to the conditional score $\nabla (\log p_t)(\cdot|\bm{y})$ where  $p_t(\cdot|\bm{y})  \triangleq p_{\bm{W}_t|\bm{Y}=\bm{y}}(\cdot)$, which, thanks to Bayes' rule can be written as 
\begin{displaymath}
	\nabla (\log p_{\bm{W}_t|\bm{Y}=\bm{y}})(\cdot) = \nabla (\log p_t)(\cdot) + \nabla (\log p_{\bm{Y} \mid \bm{W}_t = \bm{w}_t})(\cdot) \, .
\end{displaymath}
The first term is the unconditional score and is approximated with a \gls{nn} as before.  In this work, we use the \gls{dps} method \cite{chung2023diffusion} and approximate the second term by\footnote{The subscript $\bm{W}_t$ was added on $\nabla$ to specify the variable of differentiation.} 
\begin{equation} \label{eq:dps_approx}
	\nabla_{\bm{w}_t} (\log p_{\bm{Y} \mid \bm{W} = \bm{w}_t})(\bm{y}) \approx \nabla_{\bm{w}_t} (\log p_{\bm{Y} \mid \bm{W} = \hat{\bm{w}}_0(\bm{w}_t)})(\bm{y})
\end{equation}
which is the gradient of the log-likelihood  $\log p_{\bm{Y}|\bm{W}}$ of the forward model defined in \eqref{eq:poisson_disc} and \eqref{eq:forward_disc} with $\bm{W}=\hat{\bm{w}}_0(\bm{w}_t)$ ($\bm{W}$ being $\bm{X}$ or $\bm{Z}$). In this work we approximated $\log p_{\bm{Y}|\bm{W}}$ with a (negative) \gls{wls} data fidelity term (this approximation will be later used in \eqref{eq:pwls}). $\hat{\bm{w}}_0(\bm{w}_t) \triangleq \mathbb{E}_{\bm{W}_0|\bm{W}_t = \bm{w}_t}[\bm{W}_0]$  is an approximation of a noise-free image from a diffused image $\bm{w}_t$ using Tweedie's formula. The approximated conditional score is then plugged into the reverse \gls{sde} \eqref{reverse_sde} in order to generate a sample from $p_{\bm{W}|\bm{Y}=\bm{y}}$.

Using the \gls{dps} method, we implemented (i) \gls{tdps} to sample $\bm{W} = \bm{X}$ from $p_{\bm{X}|\bm{Y}=\bm{y}}$, followed by $\mathcal{F}^{\dag}$ to obtain the material images $\bm{z}$ and (ii)  \gls{odps} to directly sample $\bm{W} = \bm{Z}$ from $p_{\bm{Z}|\bm{Y}=\bm{y}}$.

\subsubsection{Implementation}\label{sec:impl} 

A U-Net architecture \cite{u_net} is used \cite{ho2020denoising, song2021scorebased} with residual blocks \cite{Resnet} that also contain time embeddings \cite{vaswani2017attention} from the \gls{dm}. Training was performed with \texttt{Adam} optimizer from PyTorch. Additionally, since Tweedie's formula necessitates one activation of the \gls{nn}, the gradient of equation \eqref{eq:dps_approx} is computed using automatic differentiation via the PyTorch function \texttt{torch.autograd}.

%% file: content/results.tex
\section{Experiments and Results} \label{section:res}

All the reconstructions methods and simulations were implemented in Python. The models were implemented and trained in Pytorch, while we used TorchRadon \cite{torch_radon} for the \gls{2d} \gls{ct} fan-beam projector. 

\subsection{Data Preparation}

\begin{figure}[!h] 
    \centering
    \begin{subfigure}{0.4\textwidth} 
        \centering
        \includegraphics[width=0.8\textwidth]{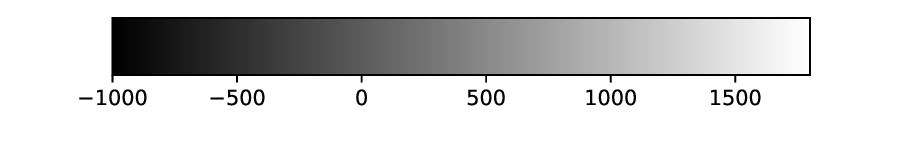}
    \end{subfigure}

    \centering
    \begin{subfigure}{0.15\textwidth}
        \includegraphics[width=\textwidth]{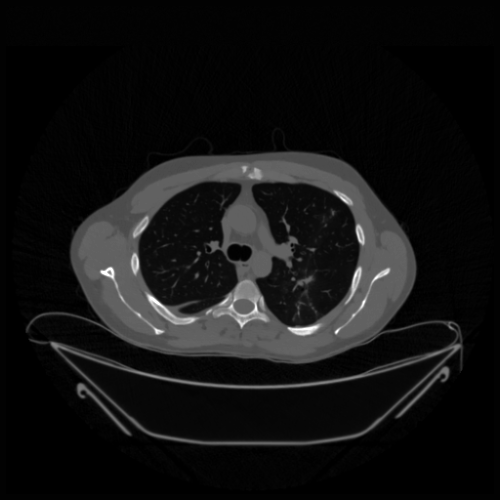}
        \subcaption{$40$ keV}
    \end{subfigure}%
    \vspace{0.1cm}
    \begin{subfigure}{0.15\textwidth}
        \includegraphics[width=\textwidth]{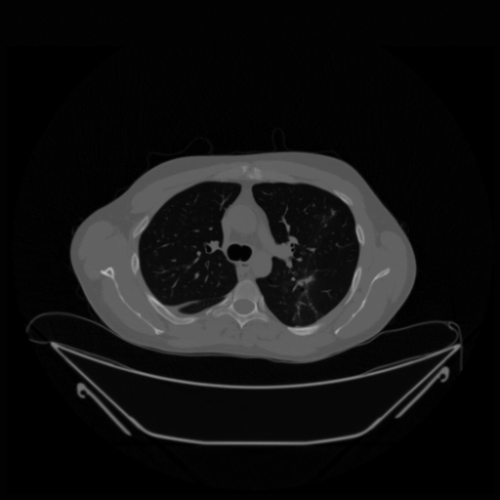}
        \subcaption{$80$ keV}
    \end{subfigure}%
    \vspace{0.1cm}
    \begin{subfigure}{0.15\textwidth}
        \includegraphics[width=\textwidth]{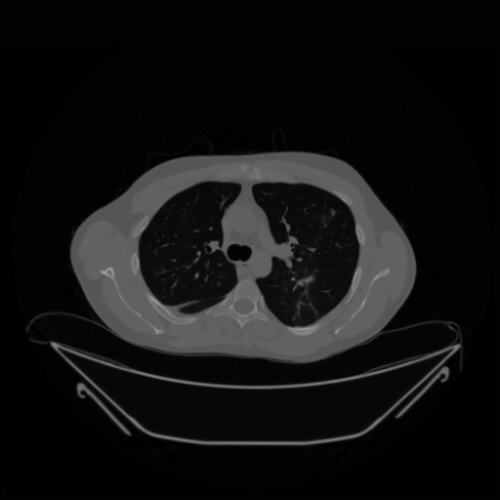}
        \subcaption{$120$ keV}
    \end{subfigure}
    
    \caption{Example of a three-energy bins attenuation image. Images are displayed in Hounsfield unit.}
    \label{fig:example_attenuation}
\end{figure}

We consider $N = 2$ materials: soft tissues and  bones. The mass attenuation coefficients  of those materials used to define $\mathcal{F}$ can be found on the \gls{nist} database \cite{nist}. We discretized both forward and reverse \glspl{sde} using Euler-Maruyama scheme with $T=1{,}000$ steps. The dataset used for this experiment consists of $11$ three-dimensional chest \glspl{ct}  at $K=3$ energy bins  ($40$, $80$ and $120$ keV), cf. Figure~\ref{fig:example_attenuation},  from Poitiers University Hospital, France, split into a training (9 patients), a validation (1 patient) and a test set (1 patient). Each slice is a  $512\times 512$ matrix  with 1-mm pixel size. Material images for training were obtained by applying $\mathcal{F}^\dag$ onto the attenuation images. An example of reference material images used to test the methods and to compute the metrics, i.e., \gls{ssim} and \gls{psnr}, is shown in Figure~\ref{fig:one_slice} (first row).

Simulated data were generated from the material images $\bm{z}$ using the forward  model described in \eqref{eq:poisson_disc} and \eqref{eq:forward_disc} with source intensity set as $\bar{h}_{i,k} = 5{,}000$ (low dose) and $\bar{h}_{i,k} = 10{,}000$. We used the ASTRA toolbox \cite{van2016fast} to implement a fan-beam projection geometry with a $120$-degree angle, incorporating 750 detectors, each with a width of 1.2~mm, with source-to-origin and origin-to-detector distances both equal to 600~mm.

\subsection{Reconstruction Methods}

We compared \gls{tdps} and \gls{odps} with two two-step image domain material decomposition methods consisting of applying $\mathcal{F}^\dag$ to the multi-energy image $\hat{\bm{x}}$ obtained by \gls{pwls}, i.e.,
\begin{align}
	& \hat{\bm{x}}  \in \argmin_{\bm{x} = \{\bm{x}_k\}}  \,   \frac{1}{2} \sum_{k=1}^K\left\| \bm{\mathcal{A}}(\bm{x}_k) - \bm{b}_k \right\|_{\bm{\Lambda}_k}^2    +    \beta R(\bm{x}) \label{eq:pwls} \\
	& \hat{\bm{z}} = {}  \mathcal{F}^\dag (\hat{\bm{x}}) \nonumber
\end{align}
where the first term in \eqref{eq:pwls} is a \gls{wls} approximation of the negative log-likelihood $\bm{x}\mapsto - \log p_{\bm{Y}|\bm{X} = \bm{x}}(\bm{y})$,  $\bm{b}_k = [b_{1,k},\dots,b_{I,k}]\in\mathbb{R}^I$ with $b_{i,k} = \log \bar{h}_{i,k}/y_{i,k}$ (with $y_{i,k}>0$), $\bm{\Lambda}_k \in\mathbb{R}_+^{I\times I}$ is a diagonal matrix of statistical weights, $R$ is a regularizer applied to the images individually or synergistically and $\beta>0$. We first considered standard \gls{wls} reconstruction ($\beta=0$) then the \gls{dtv} regularizer \cite{Synergistic_DTV} which enforces structural similarities with a clean reference  image (obtained by reconstructing from the non-binned data) and promotes the sparisty of the gradient. We used a separable quadratic surrogate algorithm \cite{Elbakri2002} for \gls{wls}  while we used a Chambolle-Pock algorithm \cite{chambolle2011first} for \gls{dtv}. 
The metrics for evaluation, i.e., \gls{ssim} and  \gls{psnr}, were computed with the Python library \texttt{skimage.metrics}.

\subsection{Results}

Figure \ref{fig:one_slice} shows material decomposition results for one slice with a $\bar{h}_{i,k}=10{,}000$ X-ray photon flux.   \Gls{wls}-reconstructed images appear noisy while the noise is somehow controlled on the \gls{dtv}-reconstructed images,  but some features  appear over-smoothed, especially in the magnified areas (spice and lungs).  \Gls{tdps} and \Gls{odps} images do not appear to suffer from noise amplification or over-smoothing, as all features seem to be preserved.

\input{./results/bigfigure2}

\begin{figure}[!h]
	\centering
    \begin{subfigure}{0.5\textwidth}
        \includegraphics[width=\textwidth]{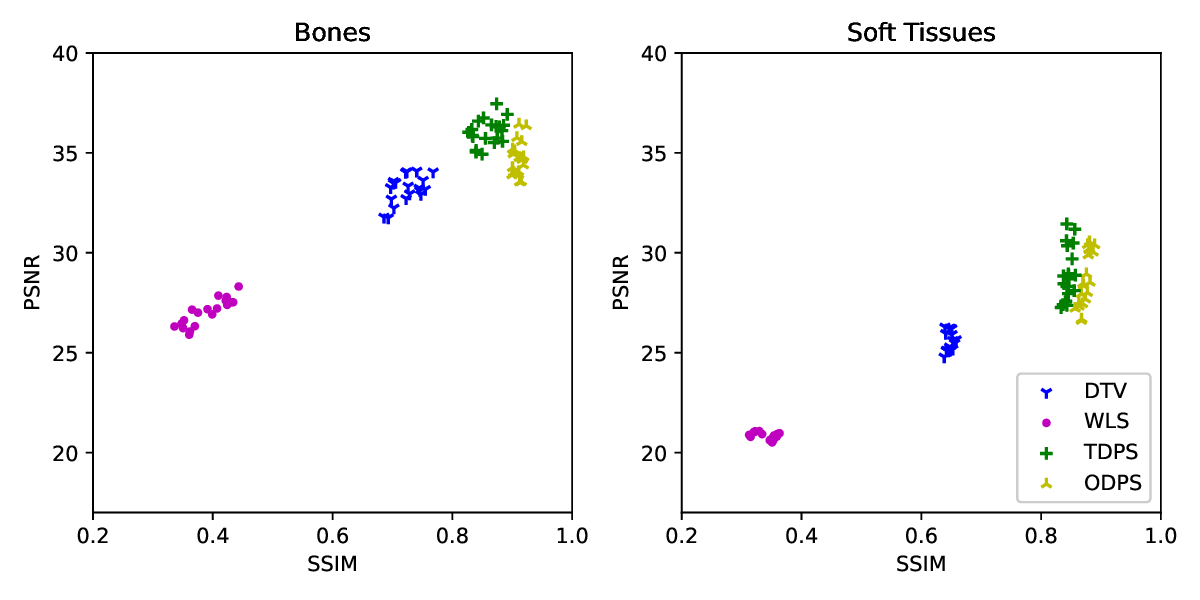}
        \vspace{-0.8cm}
        \subcaption{$\bar{h}_{i,k}=10,000$.}
        \label{fig:metrics_md1}
    \end{subfigure}
    \vspace{0.5cm}
    \begin{subfigure}{0.5\textwidth}     
        \includegraphics[width=\textwidth]{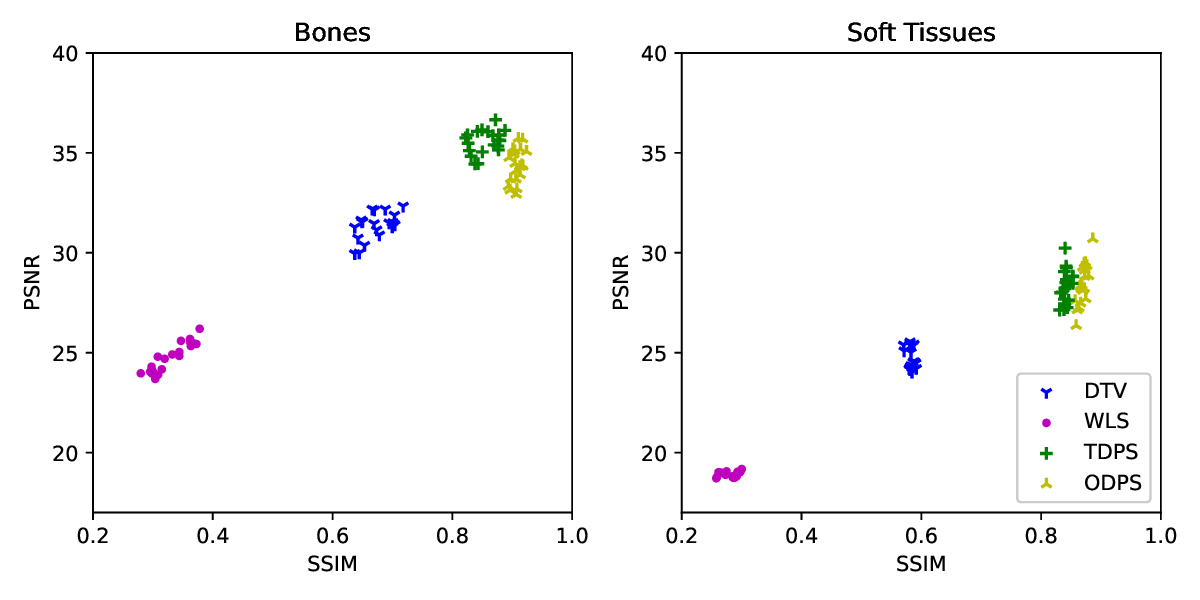}
        \vspace{-0.8cm}
        \subcaption{$\bar{h}_{i,k}=5,000$.}
        \label{fig:metrics_md1}
    \end{subfigure}
    \caption{\gls{psnr} and \gls{ssim} of $15$ slices reconstructed from $\bar{h}_{i,k}=10,000$ and $\bar{h}_{i,k}=5,000$ photon flux.}
    \label{fig:multi_res}
\end{figure}

Figure~\ref{fig:multi_res} presents the \gls{psnr} and \gls{ssim} metrics of $15$ material decompositions from each of the presented methods  with $\bar{h}_{i,k}=5{,}000$ and $\bar{h}_{i,k}=10{,}000$. The metrics were computed for each material separately. The results seem to confirm the observations from Figure~\ref{fig:one_slice}. Both \gls{odps} and \gls{tdps} appear to outperform \gls{wls} and \gls{dtv}, especially for $\bar{h}_{i,k}=5{,}000$. Finally, \Gls{odps} seems to slightly outperform \gls{tdps} in terms of \gls{ssim}.

%% file: results/bigfigure2.tex
\begin{figure}[!ht] 
	
	\centering
	\settoheight{\tempdima}{\includegraphics[width=0.4\linewidth]{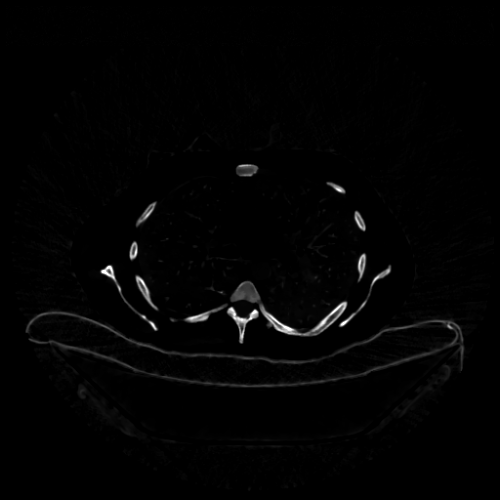}}%
	\begin{tabular}{@{\hspace{-0.7cm}}c@{\hspace{-0.01cm}}c@{\hspace{-0.6cm}}c}
		Bones &  Soft tissues &   \\	
		\vspace{-0.5cm} \\
		\rowname{\normalsize Reference}
		\begin{tikzpicture}
			\begin{scope}[spy using outlines={rectangle,yellow,magnification=2,size=8mm,connect spies}]
				\node{\includegraphics[height=\tempdima]{Res/GT_0.png}};
				\spy on (-0.05,-0.5) in node [left] at (-0.9,1.2);
			\end{scope}
		\end{tikzpicture} 
		&
		\begin{tikzpicture}
			\begin{scope}[spy using outlines={rectangle,yellow,magnification=2,size=8mm,connect spies}]
				\node {\includegraphics[height=\tempdima]{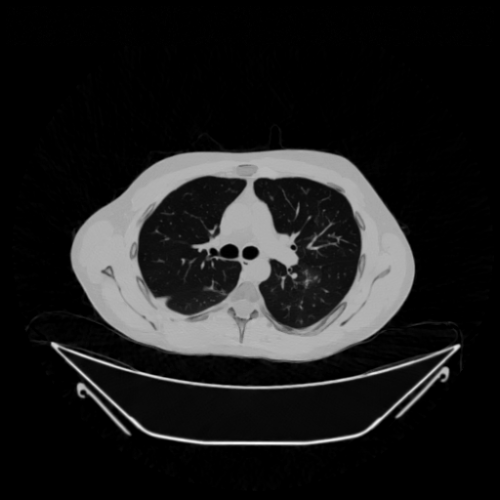}};
				\spy on (0.5,0.15) in node [left] at (1.6,1.2); 
                \spy on (-0.05,-0.5) in node [left] at (-0.9,1.2);
			\end{scope}
		\end{tikzpicture} 
	\vspace{-0.1cm}
	\\ 
		\rowname{\normalsize WLS}
		\begin{tikzpicture}
			\begin{scope}[spy using outlines={rectangle,yellow,magnification=2,size=8mm,connect spies}]
				\node {
					\begin{overpic}[height=\tempdima]{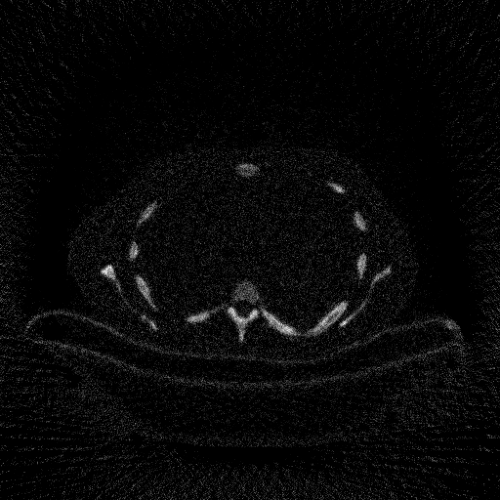}
						\put(30,15){\footnotesize\textbf{\color{white}{PSNR: 27.40}}}
						\put(30,5){\footnotesize \textbf{\color{white}{SSIM: 0.424}}}
					\end{overpic}
				};
				\spy on (-0.05,-0.5) in node [left] at (-0.9,1.2);
			\end{scope}
		\end{tikzpicture}  
	  &
	  \begin{tikzpicture}
	  	\begin{scope}[spy using outlines={rectangle,yellow,magnification=2,size=8mm,connect spies}]
	  		\node {
	  			\begin{overpic}[height=\tempdima]{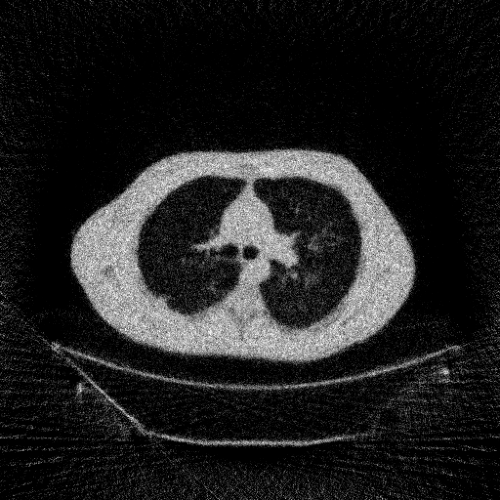}
	  				\put(30,15){\footnotesize\textbf{\color{white}{PSNR: 20.64}}}
	  				\put(30,5){\footnotesize\textbf{\color{white}{SSIM: 0.353}}}
	  			\end{overpic}
	  		};
				\spy on (0.5,0.15) in node [left] at (1.6,1.2); 
                \spy on (-0.05,-0.5) in node [left] at (-0.9,1.2);
	  	\end{scope}
	  \end{tikzpicture}  
  \vspace{-0.1cm}
  \\
  	\rowname{\normalsize DTV}
  	\begin{tikzpicture}
  		\begin{scope}[spy using outlines={rectangle,yellow,magnification=2,size=8mm,connect spies}]
  			\node {
  				\begin{overpic}[height=\tempdima]{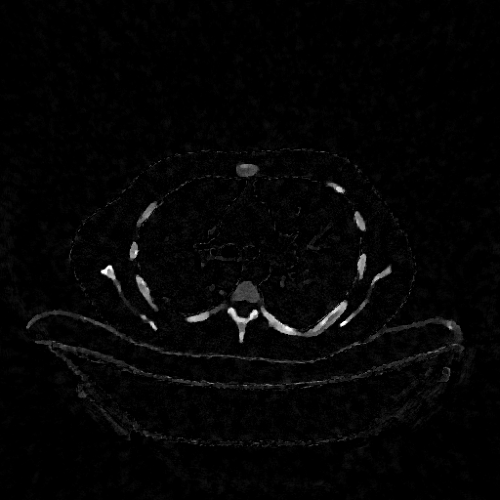}
  					\put(30,15){\footnotesize\textbf{\color{white}{PSNR: 33.12}}}
  					\put(30,5){\footnotesize\textbf{\color{white}{SSIM: 0.746}}}
  				\end{overpic}
  			};
  			\spy on (-0.05,-0.5) in node [left] at (-0.9,1.2); 
  		\end{scope}
  	\end{tikzpicture}
  &
  \begin{tikzpicture}
  	\begin{scope}[spy using outlines={rectangle,yellow,magnification=2,size=8mm,connect spies}]
  		\node {
  			\begin{overpic}[height=\tempdima]{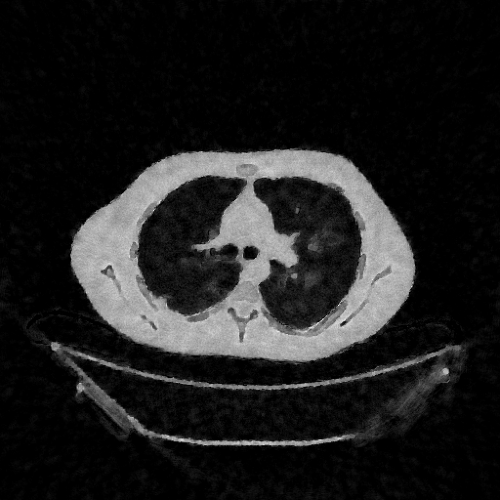}
  				\put(30,15){\footnotesize\textbf{\color{white}{PSNR: 25.15}}}
  				\put(30,5){\footnotesize\textbf{\color{white}{SSIM: 0.648}}}
  			\end{overpic}
  		};
				\spy on (0.5,0.15) in node [left] at (1.6,1.2); 
                \spy on (-0.05,-0.5) in node [left] at (-0.9,1.2);
  	\end{scope}
  \end{tikzpicture}  
	\vspace{-0.1cm}
\\
	\rowname{\normalsize TDPS}
	\begin{tikzpicture}
		\begin{scope}[spy using outlines={rectangle,yellow,magnification=2,size=8mm,connect spies}]
			\node {
				\begin{overpic}[height=\tempdima]{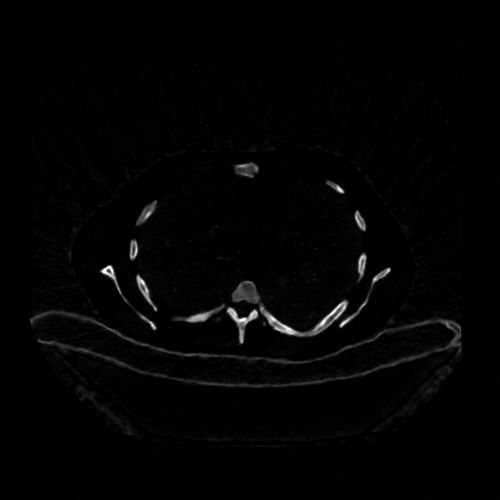}
					\put(30,15){\footnotesize\textbf{\color{white}{PSNR: 35.74}}}
					\put(30,5){\footnotesize\textbf{\color{white}{SSIM: 0.875}}}
				\end{overpic}
			};
			\spy on (-0.05,-0.5) in node [left] at (-0.9,1.2); 
		\end{scope}
	\end{tikzpicture}  
	&
	\begin{tikzpicture}
		\begin{scope}[spy using outlines={rectangle,yellow,magnification=2,size=8mm,connect spies}]
			\node {
				\begin{overpic}[height=\tempdima]{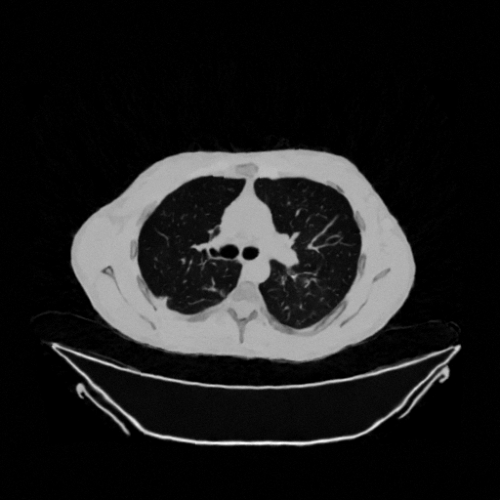}
					\put(30,15){\footnotesize\textbf{\color{white}{PSNR: 28.31}}}
					\put(30,5){\footnotesize\textbf{\color{white}{SSIM: 0.843}}}
				\end{overpic}
			};
				\spy on (0.5,0.15) in node [left] at (1.6,1.2); 
                \spy on (-0.05,-0.5) in node [left] at (-0.9,1.2);
		\end{scope}
	\end{tikzpicture} 
	\vspace{-0.1cm} 
	\\
	
	\rowname{\normalsize ODPS}
	\begin{tikzpicture}
		\begin{scope}[spy using outlines={rectangle,yellow,magnification=2,size=8mm,connect spies}]
			\node {
				\begin{overpic}[height=\tempdima]{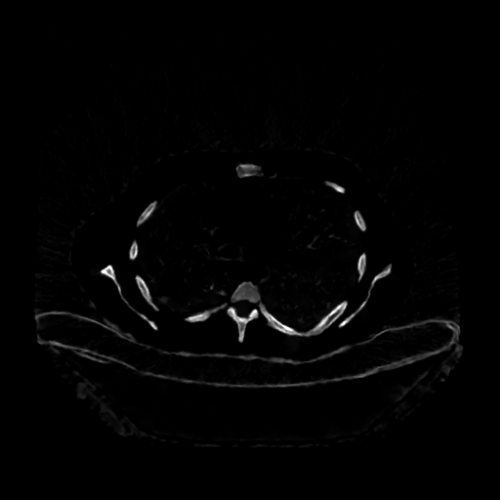}
					\put(30,15){\footnotesize\textbf{\color{white}{PSNR: 36.64}}}
					\put(30,5){\footnotesize\textbf{\color{white}{SSIM: 0.918}}}
				\end{overpic}
			};
			\spy on (-0.05,-0.5) in node [left] at (-0.9,1.2);   
		\end{scope}
	\end{tikzpicture}  
	&
	\begin{tikzpicture}
		\begin{scope}[spy using outlines={rectangle,yellow,magnification=2,size=8mm,connect spies}]
			\node {
				\begin{overpic}[height=\tempdima]{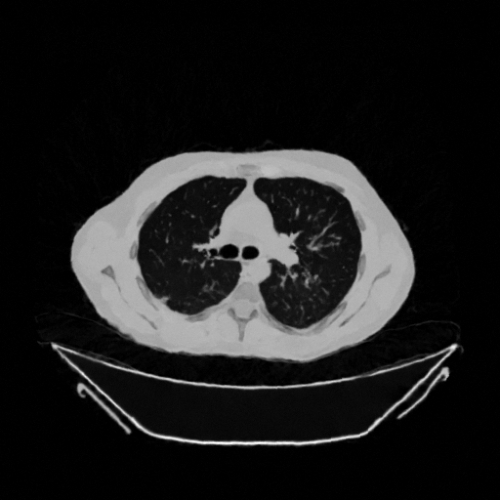}
					\put(30,15){\footnotesize\textbf{\color{white}{PSNR: 28.87}}}
					\put(30,5){\footnotesize\textbf{\color{white}{SSIM: 0.858}}}
				\end{overpic}
			};
				\spy on (0.5,0.15) in node [left] at (1.6,1.2); 
                \spy on (-0.05,-0.5) in node [left] at (-0.9,1.2);
		\end{scope}
	\end{tikzpicture}  
		
	\end{tabular}	

	\caption{Decomposition results on one slice with X-ray photon flux set to $\bar{h}_{i,k}=10{,}000$ for all energy bins. The first row are the reference material images used to compute \gls{ssim} and \gls{psnr}.}\label{fig:res_images}
\label{fig:one_slice}
\end{figure}

%% file: content/conclusion.tex
\section{Discussion and Conclusion}\label{section:conc}

Designing a prior (or a regularization function) is a central question for solving inverse problems and generative models may offer an elegant and effective solution to this. We proposed two methods for material decomposition using a reversed \gls{dm} implemented with a trained \glspl{nn} as a prior, for both \gls{tdps} and \gls{odps}. 
Both methods give promising results as compared with state-of-the-art techniques, especially \gls{tdps}.  
W expect to obtain better results with \gls{odps} in the future though fine tuning and better training (more epochs, larger database). In fact, unbeknownst to us, similar research by X. Jiang et al. ~\cite{jiang2024ct} was carried out at the time of preparing this paper. They used a ``jumpstart'' method which consists in starting the conditional diffusion on a scout decomposition and refining the computation of the gradient \eqref{eq:dps_approx}.